# The Proudman Resonance Modified: Forced Waves in Shallow Water – the Suppression of Resonance by Nonlinearity

Chunyan Li

Department of Oceanography and Coastal Sciences, Louisiana State University

May 2023

Corresponding: cli@lsu.edu

**Abstract.**

Storms and severe weather are often associated with atmospheric low-pressure systems. A moving low atmospheric pressure system produces storm surge that is dependent on the wave Froude number ($F$) or the ratio between the speed of the storm and the shallow water wave propagation speed. Proudman's resonance for such a problem goes to infinity at $F = 1$. This work revisits the problem by presenting a nonlinear exact solution. The first order approximation goes back to Proudman's solution. The second order approximation provides a finite solution including at resonance. The solution shows that the nonlinear advection can significantly suppress the resonance. At resonance, although the solution is finite, it bifurcates into two branches, a condition linked to chaos in theoretical fluid dynamics and many nonlinear systems. Although Proudman's solution does not provide any limit to the atmospheric pressure system, it is not without any condition. The solution exists for all low-pressure systems but has only limited validity for a moving atmospheric high-pressure system induced forced waves: at low moving speed, the solution under a high-pressure system can exist for a range of high-pressure variability but quickly become invalid as the system's speed increases. The closer $F$ is to 1 (resonance), the easier for the solution to be invalidated for a high-pressure system. This study also found that Proudman's solution is most accurate only under subcritical conditions with small low-pressure variability and has much larger error under supercritical conditions.

## 1. Introduction

The intensities of tropical cyclones are often classified by the Simpson-Saffer Scale (SSS) which is uniquely defined (Saffir, 1973; Simpson, 1974) by a single parameter: the maximum sustained wind speed (MSWS). A Category 1 hurricane is defined as having the MSWS of 119-153 km/hr.; and the Categories 2,3,4, and 5 are defined as those having MSWS of 154-177 (Cat 2), 178-208 (Cat 3), 209-251 (Cat 4), and greater than 252 km/hr. (Cat 5), respectively. This classification is easy to apply as it should be, for the public to understand and use in practice. However, it is oversimplified by ignoring influences from several other factors known for years.

As some examples, as early as 1950s, it was found that the central air pressure could explain 50% variability of the maximum hurricane storm surge although no clear correlation was found between the storm surge and radius of maximum wind at that time (Conner et al., 1957; Harris, 1959) before the era of satellite observations (Hasler and Morris, 1986; Folmer et al., 2012; Han et al., 2017), weather RADARs (Yang et al., 2019; Shen et al., 2020), and before the hurricane hunting airplane measurements have accumulated sufficient data (Aberson et al., 2017; Zhu et al., 2023). Jelesnianski (1966) conducted numerical experiments, using linearized transport equations without bottom friction, on the effects of the maximum wind speed, radius of maximum wind (RMW), direction of motion, and the moving speed of hurricane. The study confirmed some of the results of the 1950s (Conner et al., 1957; Harris, 1959) and found that the moving speed of the storm plays a role in the tropical cyclone's sensitivity to the initial location of the storm, particularly for slow moving storms. Irish et al. (2008) studied the effect of RMW and demonstrated that it could have significant effect on storm surges, particularly over mildly sloping coasts.

In Peng et al. (2004), a simulation of 10 hypothetical hurricanes allowed a conclusion that a hurricane with lower minimum central pressure or larger RMW would generate greater storm surge and a larger inundation area. The minimum central pressure had a greater influence than the RMW. In addition, a hurricane with slower moving speed would produce higher storm surge and larger inundation area. Using Hurricane Hugo of 1989 as a prototype, Peng et al. (2006a) experimented with hypothetical hurricanes from different angles or along different tracks and concluded that both the storm surge and inundation area are sensitive to the storm track. Peng et al. (2006b) further proceeded with more parameter space analysis for the effects of storm moving speed, RMW, and inflow angle.

In a numerical model experiment (Rego and Li, 2009), Hurricane Rita induced storm surge was used as a control and several characteristic parameters of the hurricane were varied to study the effect of their impact to storm surge flood volume with 21 simulations using the Finite Volume Community Ocean Model (FVCOM). In that study, several parameters were examined: the maximum wind speed, RMW, tidal phase at landfall time, tidal amplitude, moving speed of the hurricane, and the inflow angles of the cyclonic wind field. The wind intensity and RMW were found to have significant impact on the flood volume. Landfall timing had some but limited impact on the flood volume and peak storm surge. In addition, the forward translation speed of the hurricane also had a significant impact on the flood volume and peak storm surge values. Varying the forward speed of the hurricane while keeping other parameters identical resulted in the flood volume fluctuating with a magnitude equivalent to an upgrade or downgrade of about 1 category on the Saffir-Simpson scale, indicating the importance of the hurricane's translational speed.

In Zhang and Li (2019), more than 1200 simulations were done using FVCOM for an idealized numerical experiment to determine the variability of the storm surge in response to the hurricane moving speed and approach angle. It was found that the moving speed and approach angle both had significant influence on the maximum storm surge.

These studies provided useful databases and conclusions in understanding the variability of storm surges based on the parameter space variables. More specifically, the above examples of study highlight the importance of several factors other than the maximum wind speed of the tropical cyclone in determining the storm surge, inundation area, or flood volume. Among these factors, the tropical cyclone's forward speed is of most interest here. This is because of the phenomenon called the Proudman (1929) resonance. This is relevant not only to hurricane storm surges but also to meteotsunami which is weather induced wave and surge phenomenon with some overlap in spectrum with the hurricane storm surges (e.g., Šepić et al., 2009; Sheremet et. al., 2016; Pellikka et. al., 2022).

With several variables and the complications demonstrated by these studies, our focus is on the basic mechanisms, particularly pertinent to the Proudman resonance because in theory the storm surge goes to infinity at resonance (Proudman, 1953). In this paper, we will revisit the original Proudman resonance problem by solving the nonlinear equations exactly. The solution gives a finite function at Proudman resonance. Discussion will be made on the behavior of the solution including when resonance is approached.

## 2. The mathematical problem

Storms and severe weather are often associated with atmospheric low-pressure systems. Examples include tropical cyclones, extra-tropical cyclones, atmospheric cold fronts, squall limes, super cell thunderstorms, mesoscale convective systems (MCS), and mesoscale convective complexes (MCC). The Proudman problem is one relevant to the simplest ocean response to a moving storm and can be described as this: in a horizontally boundless shallow sea with a constant water depth $h$, extending infinitely in both positive and negative x-directions, a steady atmospheric low-pressure system moves at a constant speed $U$ in the x-direction, producing a forced wave moving with the atmospheric system, the steady state linear solution is (Proudman, 1929, 1953):

$$\zeta = \frac{\tilde{\zeta}}{1 - \frac{U^2}{gh}} \tag{1}$$

in which $g$ is the gravitational acceleration at the Earth's surface, $\zeta$ is the water level change, $\tilde{\zeta}$ is the hydrostatic water level change due to the atmospheric pressure change from the environmental air pressure, i.e.

$$\tilde{\zeta} = -\frac{\Delta p}{\rho g}$$

where $\Delta p$ is the atmospheric pressure difference from the environmental value (outside of the weather system), and $\rho$ is the seawater density. The main parameter that determines the solution is the wave Froude number $F = U/\sqrt{gh}$, in contrast to the (particle) Froude number $F_p =$ $u/\sqrt{gh}$. Equation (1) shows that when the atmospheric system moves at the same speed as the shallow water wave propagation speed, i.e., the wave Froude number is unity or when $U = \sqrt{gh}$, the water level would go to infinity, indicating a divergent solution. In this paper, the Proudman's problem is revisited, and an exact solution is sought to examine the convergence of the solution at resonance (when $U = \sqrt{gh}$) to gain information on the impact of nonlinearity on the suppression of resonance. The governing equations for a one-dimensional problem relevant to this subject are

$$\frac{\partial \zeta}{\partial t} + \frac{\partial (h+\zeta)\bar{u}}{\partial x} = 0 \tag{2}$$

$$\frac{\partial \bar{u}}{\partial t} + \bar{u}\frac{\partial \bar{u}}{\partial x} = -g\frac{\partial (\zeta - \zeta_a)}{\partial x} \tag{3}$$

$$\zeta_a = -\frac{p_a}{\rho g} \tag{4}$$

with these conditions:

(i) $h = \text{constant} > 0$

(ii) $-\infty < x < +\infty$

(iii) $-\infty < t < +\infty$

(iv) $-\infty < U = \text{constant} < +\infty$

(v) $\zeta_a = \zeta_0(x - Ut)$ (5)

(vi) $\zeta \to 0, \bar{u} \to 0$ when $\zeta_0 \to 0$ (6)

Given the parameters $U$ and characteristic time scale $T$, we can convert the variables to non-dimensional forms

$$Z = \frac{\zeta}{h}, Z_a = \frac{\zeta_a}{h}, u = \frac{\bar{u}}{U}, X = \frac{x}{UT}, \tau = \frac{t}{T}, n = \frac{U^2}{gh} \tag{6}$$

$$\frac{\partial Z}{\partial \tau} + \frac{\partial (1+Z)u}{\partial X} = 0 \tag{7}$$

$$\frac{\partial u}{\partial \tau} + u\frac{\partial u}{\partial X} = -\frac{1}{n}\frac{\partial (Z - Z_a)}{\partial X} \tag{8}$$

The conditions are now

(I) $n \geq 0$

(II) $-\infty < X < +\infty$

(III) $-\infty < \tau < +\infty$

(IV) $Z_a = Z_0(X - \tau)$ (9)

(V) $Z \to 0, u \to 0$ when $Z_0 \to 0$ (10)

We now seek a forced wave steady state solution in non-dimensional form $Z$ and $u$ under the atmospheric disturbance $Z_a$

$$Z = Z(X - \tau) = Z(\xi) \tag{11}$$

$$u = u(X - \tau) = u(\xi) \tag{12}$$

## 3. The solution

Using (11) and (12), the continuity equation (7) leads to

$$u = \frac{Z}{(1 + Z)} \tag{13}$$

Equation (13) satisfies condition (10): $Z \to 0,$ when $Z_0 \to 0$ ($Z \to 0$ and $u \to 0$ can be satisfied if any of them is approaching to 0). Combining the momentum equation (8), we obtain the following equation

$$\left[1 - \frac{n}{(1 + Z)^3}\right] \frac{\partial Z}{\partial \xi} = \frac{\partial Z_a}{\partial \xi} \tag{14}$$

which can be integrated to

$$Z + \frac{n}{2(1 + Z)^2} = Z_a + \frac{n}{2} \tag{15}$$

A rearrangement leads to

$$Z^3 + A_1 Z^2 + A_2 Z + A_3 = 0 \tag{16}$$

in which

$$A_1 = 2 - \left(Z_a + \frac{n}{2}\right) \qquad (17)$$

$$A_2 = 1 - (2Z_a + n) \qquad (18)$$

$$A_3 = -Z_a \qquad (19)$$

Let $Z_a \to 0$, we have

$$Z(Z^2 + A_1 Z + A_2) = -A_3 \to 0$$

Thus, we have

**Lemma A.** Existence of solution: there must be a root for (16) that approaches 0 as the forcing function goes to zero, i.e., $Z \to 0$ as $Z_a \to 0$.

Assume that $Z$ and $Z'$ are two solutions satisfying (10) under the same conditions (e.g., same $U$ and forcing function $Z_a$), from (15) we can derive

$$(Z - Z')\left[1 - \frac{n(Z + Z' + 2)}{2(1 + Z)^2(1 + Z')^2}\right] \equiv 0 \qquad (20)$$

we can see that if $n \neq 1$, we must have

$$Z \equiv Z' \qquad (21)$$

Otherwise, the term within the brackets would have to be zero. But if we let $Z_a \to 0$, the term in the brackets would approach $1 - n \neq 0$, which is contradictory to the assumption so only (21) is possible. From this result, we have

**Lemma B.** Uniqueness of solution: the solution must be unique when $n \neq 1$. The condition $n \neq 1$ represents non-resonance states.

If $n = 1$, the atmospheric disturbance has a traveling speed equal to the shallow water wave propagation speed $\sqrt{gh}$, resonance occurs, and it is not guaranteed that the solution is unique for this problem. This is further discussed later in this article.

For non-resonance states, $n \neq 1$, the three roots for (16) are:

$$Z_1 = C_1 + C_2 - \frac{A_1}{3} \qquad (22)$$

$$Z_2 = \omega_1 C_1 + \omega_2 C_2 - \frac{A_1}{3} \quad (23)$$

$$Z_3 = \omega_2 C_1 + \omega_1 C_2 - \frac{A_1}{3} \quad (24)$$

in which

$$C_1 = \sqrt[3]{-\frac{q}{2} + \sqrt{\Delta}}, \qquad C_2 = \sqrt[3]{-\frac{q}{2} - \sqrt{\Delta}} \quad (25)$$

$$\omega_1 = \frac{-1 + i\sqrt{3}}{2}, \qquad \omega_2 = \frac{-1 - i\sqrt{3}}{2} \quad (26)$$

$$\Delta = \frac{q^2}{4} + \frac{p^3}{27} = n \frac{\frac{27n}{2} - 4\left(1 + \frac{n}{2} + Z_a\right)^3}{8 \times 27} \quad (27)$$

$$p = \frac{3A_2 - A_1^2}{3}, \qquad q = \frac{2A_1^3 - 9A_1A_2 + 27A_3}{27} \quad (28)$$

When $\Delta < 0$, there are three different real roots for (16), which can be expressed as

$$Z_{k+1} = 2S \cos \frac{\varphi + 2k\pi}{3} - \frac{A_1}{3}, \quad (k = 0,1,2) \quad (29)$$

in which

$$S = \sqrt[3]{r} = \sqrt{-\frac{p}{3}}, \quad \cos \varphi = -\frac{q}{2r}, r = \sqrt{-\frac{p^3}{27}}, \quad (30)$$

When $\Delta = 0$, there are three real roots, in which two of them are equal.

When $\Delta > 0$, there is one real root, and two complex roots which are complex conjugates.

If we write

$$C_0(n) = \frac{3\sqrt[3]{n}}{2} - \left(1 + \frac{n}{2}\right) \quad (31)$$

Since $n \geq 0$, it can be seen that

$$C_0(n) \leq 0 \quad (32)$$

It can also be shown that $\Delta < 0$ is equivalent to $Z_a > C_0(n)$, thus

$$\Delta < 0 \;\Leftrightarrow\; Z_a > C_0(n) \tag{33}$$

Similarly, we can see that

$$\Delta = 0 \;\Leftrightarrow\; Z_a = C_0(n) \tag{34}$$

$$\Delta > 0 \;\Leftrightarrow\; Z_a < C_0(n) \tag{35}$$

From (32)-(35), it can be deduced that

**Lemma C.** Only when $\Delta \le 0$, can we have $Z_a$ containing the real value 0, i.e.

$$0 \epsilon \{Z_a | Z_a \ge C_0(n)\} \;\text{ or }\; 0 \notin \{Z_a | Z_a < C_0(n)\} \tag{36}$$

To satisfy the condition (10), we must have $\Delta \le 0$. Therefore, if a solution of (7)-(8) exits for $\Delta > 0$, it must be an extension of a solution for $\Delta \le 0$ because when $Z_a \to 0$, the corresponding solution $Z$ is a root within $\Delta \le 0$. Now we can rewrite (29) for $\Delta < 0$ in the following format:

$$Z_k = \frac{2}{3}\left(1 + \frac{n}{2} + Z_a\right)\cos\left\{\frac{1}{3}\operatorname{acos}\left[1 - \frac{27n}{4\left(1 + \frac{n}{2} + Z_a\right)^3}\right] + \frac{2(k-1)\pi}{3}\right\} - \frac{1}{3}\left(2 - \frac{n}{2} - Z_a\right),$$

$$(k = 1,2,3) \tag{37}$$

Using (10), i.e., to satisfy $Z_k \to 0$, when $Z_0 \to 0$, we must find the k value among 0,1,2 so that the following identity is true

$$\cos\left\{\frac{1}{3}\operatorname{acos}\left[1 - \frac{27n}{4\left(1 + \frac{n}{2}\right)^3}\right] + \frac{2k\pi}{3}\right\} \equiv \frac{2 - \frac{n}{2}}{2 + n} \tag{38}$$

From Lemma A, the solution exists, and the k value must exist among k=0,1,2 to satisfy (38). To find the correct k value to have (38) to be an identity, it requires that the right-hand side (RHS) and left-hand-side (LHS) of (38) increase or decrease together with $n$. The RHS and LHS as functions of $n$ are:

$$F_k(n) = \cos\left(\frac{1}{3}\operatorname{acos} x_0 + \frac{2k\pi}{3}\right) \tag{39}$$

$$F(n) = \frac{2 - \frac{n}{2}}{2 + n} \tag{40}$$

The derivatives of (39) and (40) are, respectively,

$$\frac{dF_k(n)}{dn} = \frac{9(n-1)}{4\left(1+\frac{n}{2}\right)^4} \frac{1}{\sqrt{1-x_0^2}} \sin\left(\frac{1}{3}\operatorname{acos} x_0 + \frac{2k\pi}{3}\right) \tag{41}$$

$$\frac{dF(n)}{dn} = -\frac{3}{(2+n)^2} < 0 \tag{42}$$

i.e., the RHS is always less than 0, while the sign of the LHS is to be determined. The first equation, (41), changes sign at $n = 1$.

Since

$$0 \leq \frac{1}{3}\operatorname{acos} x_0 \leq \frac{\pi}{3} \tag{43}$$

It follows that

$$\frac{2k\pi}{3} \leq y \leq (2k+1)\frac{\pi}{3} \tag{44}$$

where

$$y = \frac{1}{3}\operatorname{acos} x_0 + \frac{2k\pi}{3} \tag{45}$$

thus

$$\sin y \geq 0 \qquad \text{when } k = 0 \text{ or } y \,\epsilon \left[0, \frac{\pi}{3}\right] \tag{46}$$

$$\sin y \geq 0 \qquad \text{when } k = 1 \text{ or } y \,\epsilon \left[\frac{2\pi}{3}, \pi\right] \tag{47}$$

$$\sin y \leq 0 \quad \text{when } k = 2 \text{ or } y \,\epsilon \left[\frac{4\pi}{3}, \frac{5\pi}{3}\right] \tag{48}$$

which lead to the results that when $k = 0$ and $k = 1$

$$\frac{dF_k(n)}{dn}\begin{cases} < 0 & (n < 1) \\ > 0 & (n > 1) \end{cases} \quad (49)$$

and when $k = 2$

$$\frac{dF_2(n)}{dn}\begin{cases} > 0 & (n < 1) \\ < 0 & (n > 1) \end{cases} \quad (50)$$

Therefore, we have the following conclusion:

**Lemma D.** The expression (38) is true when $k = 0$ if $n < 1$; or when $k = 2$ if $n > 1$.

The case when $k = 1$ must be excluded for (38) to be an identity (this can be seen by the fact that $F_1(0) = -0.5;\ F(0) = 1$). Combining the results, we have the following conclusion:

**Theorem**. The solution of (7) and (8) for the case when $n < 1$ is

$$Z = \frac{2}{3}\left(1 + \frac{n}{2} + Z_a\right)\cos\left\{\frac{1}{3}\mathrm{acos}\left[1 - \frac{27n}{4\left(1 + \frac{n}{2} + Z_a\right)^3}\right]\right\} - \frac{1}{3}\left(2 - \frac{n}{2} - Z_a\right) \quad (51)$$

when $n > 1$, the solution is

$$Z = \frac{2}{3}\left(1 + \frac{n}{2} + Z_a\right)\cos\left\{\frac{1}{3}\mathrm{acos}\left[1 - \frac{27n}{4\left(1 + \frac{n}{2} + Z_a\right)^3}\right] + \frac{4\pi}{3}\right\} - \frac{1}{3}\left(2 - \frac{n}{2} - Z_a\right) \quad (52)$$

given the conditions (I)-(V). The velocity $u$ is determined by (13).

When $n = 1$, the solution bifurcates into two branches – both (51) and (52) are solutions satisfying (7) and (8).

The solutions under different $n$ values can be verified directly by substituting the functions into the equations. This is straightforward and can be easily done but omitted here for brevity. This indirectly verifies that the identity (38) is true for $k = 0$, if $n < 1$ and $k = 2$, if $n > 1$. The identity, however, can also be verified by computations: Figure 1 shows the difference between the LHS and RHS of (38). It is basically the computational error ($\sim 10^{-15}$, except near $n = 1$ where the error is $\sim 10^{-14}$).

Now we examine the case when $\Delta > 0$. We can prove that a real solution of (7) and (8) satisfying (I)-(V) does not exist under this condition. This can be seen by examining the following:

$$A_0 = \lim_{Z_a \to C_0(n)} Z_2 = \lim_{Z_a \to C_0(n)} Z_3 = \sqrt[3]{n} - 1 \tag{53}$$

$$B_0 = \lim_{Z_a \to C_0(n)} Z_1 = -\left(1 + \frac{\sqrt[3]{n}}{2}\right) \tag{54}$$

Obviously, $A_0 \neq B_0$. Note also that $Z < 1$ means that the surface elevation is below the water bottom, which is impossible for a real solution. We therefore have this conclusion:

**Lemma E.** When $\Delta > 0$, there is no real solution for (7) and (8) satisfying conditions (I)-(V). $Z_2$ and $Z_3$ are complex conjugate of each other.

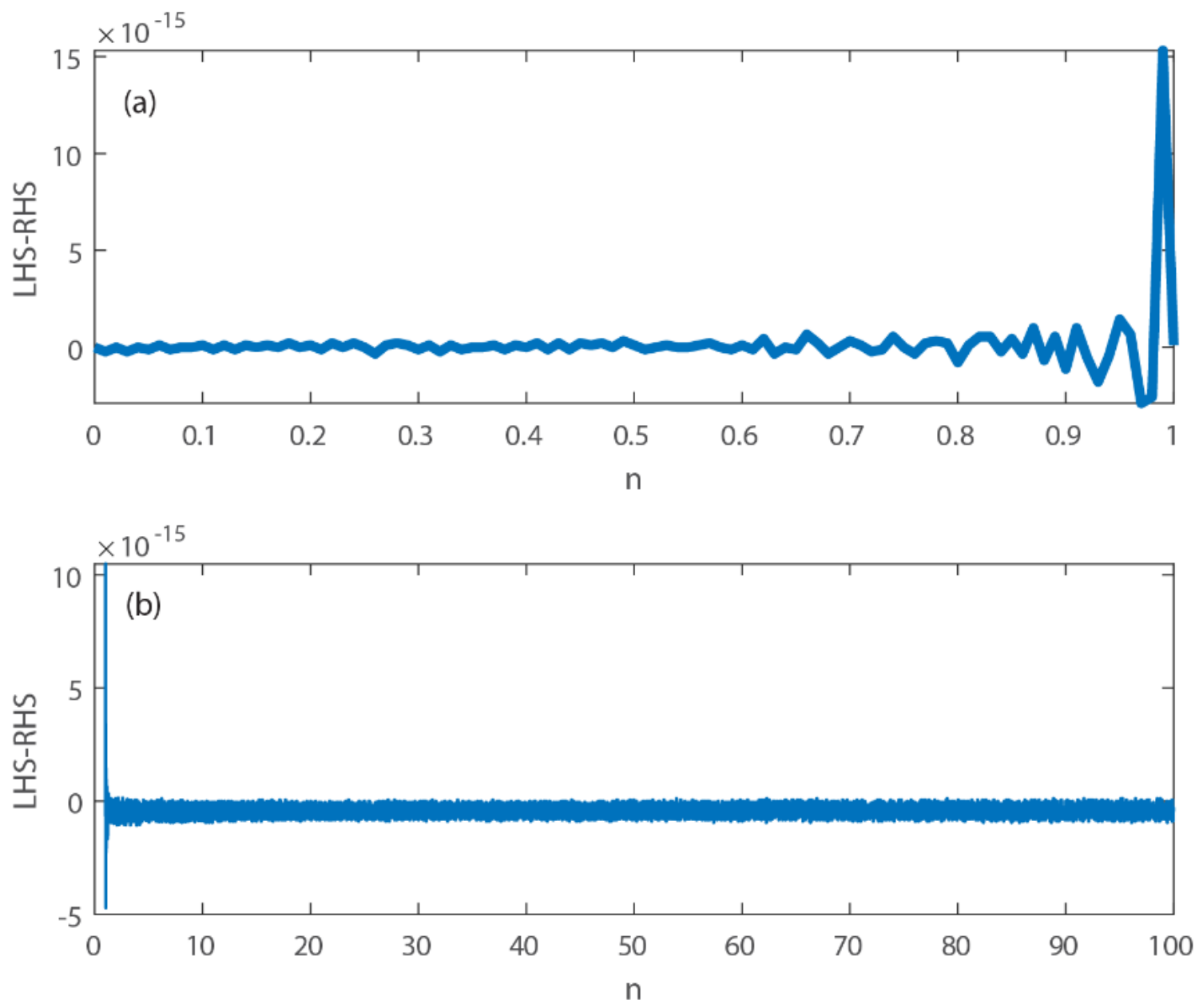

**Figure 1.** Verification of the solution. (a) For $n \leq 1$; and (b) For $n \leq 100$.

## 4. Approximations

The solutions (51) and (52) can be approximated using Taylor series expansion for the first order and second order approximations if $|Z_a| \ll 1$. The second order approximation is

$$Z \approx \frac{n - 1 \pm \sqrt{(1-n)^2 + 6nZ_a}}{3n}, \quad \left(Z_a > -\frac{(1-n)^2}{6n}\right) \tag{55}$$

Here the $\pm$ sign takes + for $n < 1$ and $-$ for $n > 1$, so that $Z \to 0$, when $Z_a \to 0$. For the resonance condition $n = 1$, the resonance solution has two branches:

$$Z \approx Z_R = \pm\sqrt{\frac{2Z_a}{3}} \qquad (Z_a > 0) \qquad (56)$$

Bifurcation is often related to chaos and fractal (e.g., Layek and Pati, 2017; Shao et al., 2023).

Under non-resonance conditions, i.e., if $n \neq 1$, the linearized approximation (first order approximation) $Z_L$ is

$$Z \approx Z_L = \frac{Z_a}{1-n} \qquad (57)$$

This is the well-known Proudman's solution, which is divergent (invalid) at resonance ($n = 1$). The conditions for the second order equation $Z_a > -\frac{(1-n)^2}{6n}$ should still be applicable. This is mostly for low-pressure systems (more discussion below). This condition, however, was not obtained from Proudman's method and solution.

## 5. Discussion and Conclusion

### *5.1. For high-pressure systems*

The original Proudman problem gives a simple linear solution without knowledge of any restrictions to the forcing function, i.e., any low- or high- or mixed low- and high-pressure systems appear to be applicable (which is not true as seen in the derivation). However, the exact solution provided here has a restriction to the forcing function. This is a combination of (33) and (34),

$$Z_a \geq C_0(n) = \frac{3\sqrt[3]{n}}{2} - \left(1 + \frac{n}{2}\right) \qquad (58)$$

From (32), $C_0(n)$ is 0 when $n = 1$ and $C_0(n) < 0$ otherwise. This means that the solutions for $n < 1$, i.e., (51) and for $n > 1$, i.e., (52) are valid for all low-pressure systems ($Z_a > 0$). For high-pressure systems, there is no solution at resonance ($n = 1$). At non-resonance condition ($n \neq 1$), a high-pressure system may allow such a forced wave solution in the form of (51) or (52) but the pressure value must satisfy (58). Figure 2 shows the domain of validity of the solution under high-pressure systems. It is determined by the Δ function of equation (27). The shaded area has positive Δ values. Only negative Δ values allow the steady state forced wave solution to exist: the shaded area is a region of no-solution. The solution is always valid for low-pressure systems and further away from the resonance condition, the solution is more forgiving for high-pressure systems.

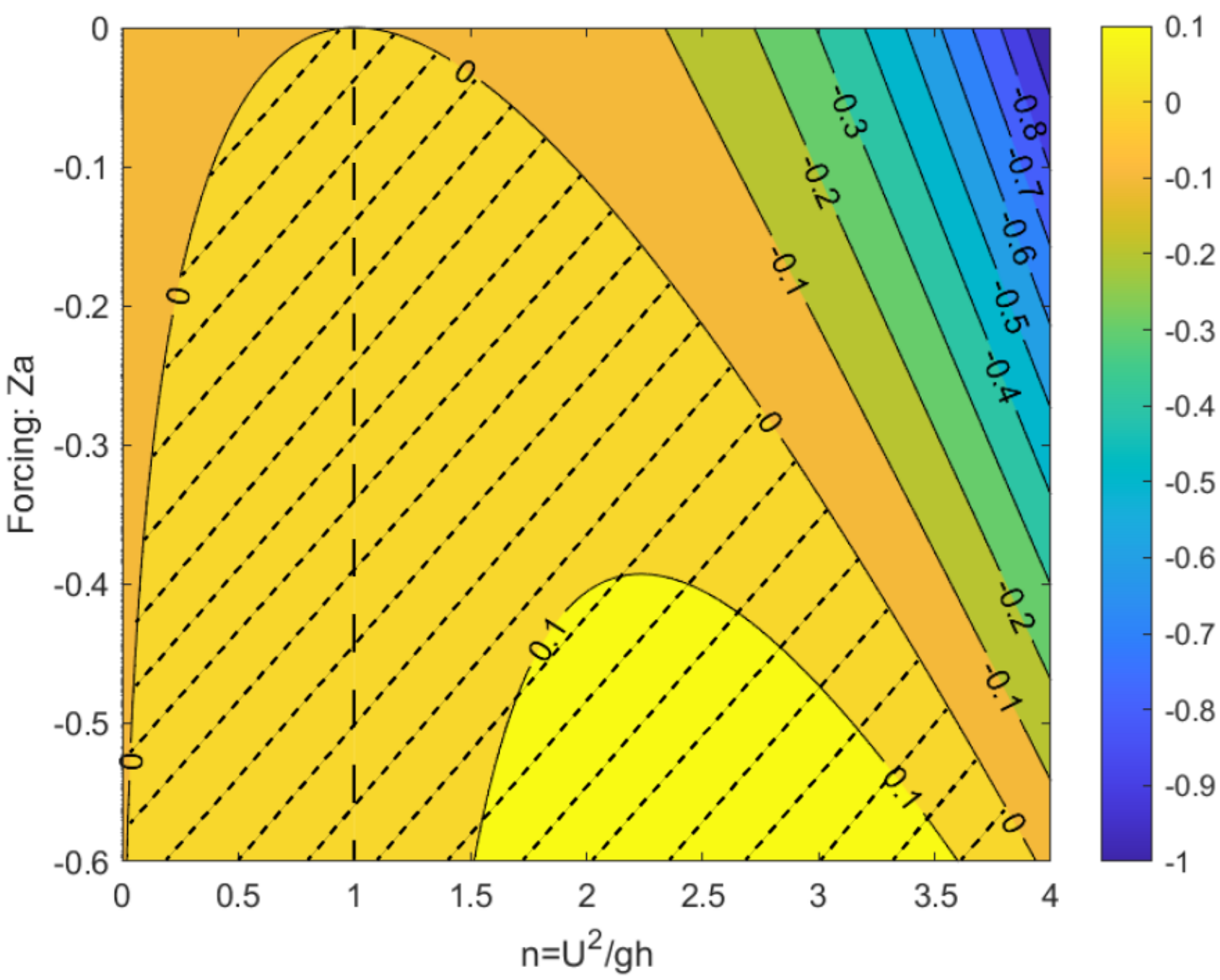

**Figure 2.** Domain of validity of the solution under high-pressure. The contour values are for Δ determined by equation (27).

*5.2. Comparison*

Figure 2 provides a comparison among the exact solution equation (51), the nonlinear approximate solution equation (55), and the Proudman's linear solution equation (57) for the subcritical scenarios ($n < 1$). When $n$ is small (e.g., $n = 0.1$), the three are very close. As $n$ increases, the linear solution departs from the exact and approximate solutions and the greater the forcing function $Z_a$, the larger the error will be. On the other hand, larger $Z_a$ will only lead to slightly increased difference between the exact and approximate solutions. Check the lower right panel of Figure 2, which is when $n = 0.9$, the Proudman solution has such a large error that the near resonance solution is essentially invalid, while the approximate solution stays close to the exact solution. Proudman's solution is far off near resonance, particularly for large forcing $Z_a$. Apparently, nonlinearity suppresses the resonance. For the supercritical scenarios, the difference is even more remarkable. Figures 3 and 4 show the comparisons for supercritical and subcritical conditions, respectively. Unless the forcing function $Z_a$ is very small, the Proudman's solution departs from the exact solution quickly as $Z_a$ increases. Again, the approximate solution stays very close to the exact solution.

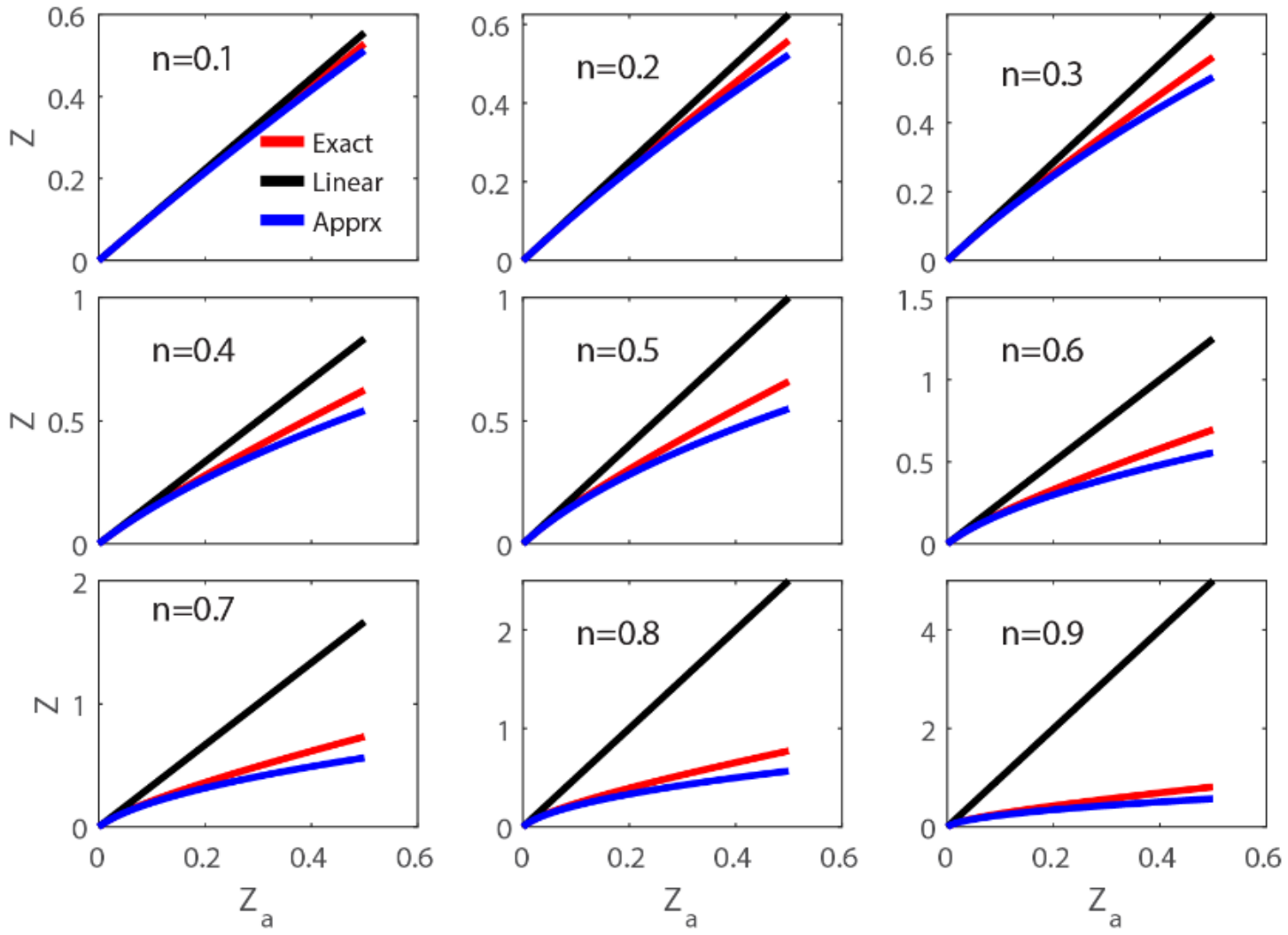

**Figure 3.** The exact, approximate, and Proudman solutions under subcritical conditions, i.e., when $n < 1$ or $F < 1$ .

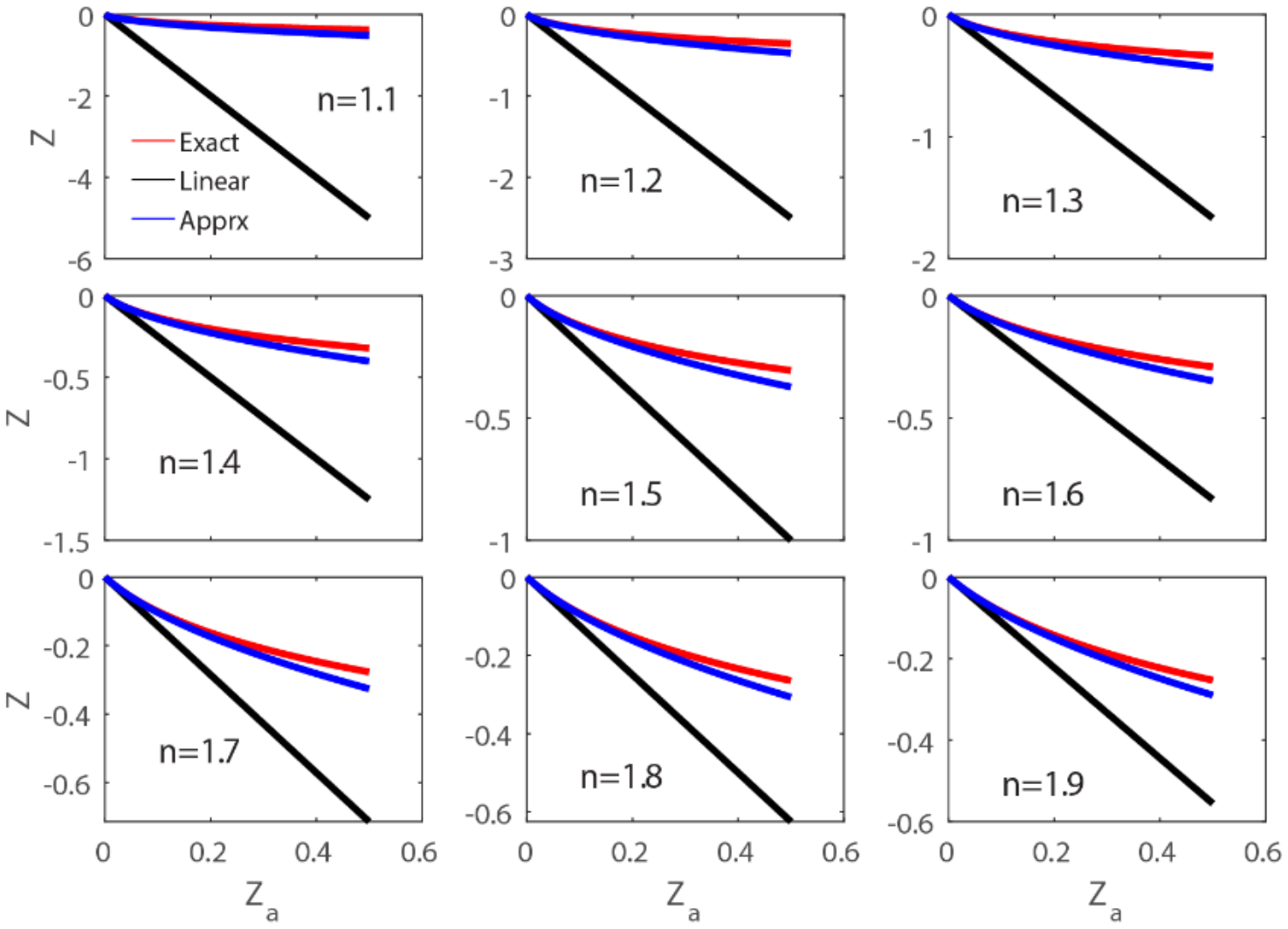

**Figure 4.** The exact, approximate, and Proudman solutions under supercritical conditions, i.e., when $n > 1$ or $F > 1$.

*5.3. Conclusions*

This short article revisits Proudman's resonance problem for a moving atmospheric pressure system and provides the derivation and verification of an exact solution – equations (51) and (52) or the subcritical and supercritical conditions, respectively. The condition that allows the exact solution is also given – equation (58). The single major parameter that determines the characteristics of the solution is the non-dimensional number, the wave Froude number

$$F = \frac{U}{\sqrt{gh}} = \sqrt{n} \tag{59}$$

The condition (58) is applicable not only to the exact solution, but also to the original Proudman's linear solution – equation (1) in dimensional form or (57) in non-dimensional form. The exact solution is finite at resonance (when $n = 1$ or wave Froude number $F = 1$). Condition (58) says that this kind of steady state moving forced wave exists for all low-pressure atmospheric moving systems. For moving high-pressure systems, however, the condition is not so forgiving – it depends on the wave Froude number. When the wave Froude number is unity, there is no solution permitted for any high-pressure system. As the wave Froude number changes away from unity, the solution for high-pressure can exist and the range of allowed high-pressure change increases as the wave Froude number moves away from 1. At resonance, when $F = 1$, the exact solution bifurcates into two branches, but both remain finite. In fact, the exact solution is always finite.

This work also provides the first and second order approximations of the exact solution using Taylor series expansion. The first order approximation is Proudman's linear solution, while the second order approximation is a much improved one in the sense that it remains finite at resonance, it bifurcates at resonance, and it has much smaller error. Comparisons among the exact, linear, and second order approximation solutions showed that Produman's solution is only meaningful for 1) very small air-pressure changes, 2) subcritical conditions ($n < 1$ or $F < 1$), and 3) when Froude number is very small ($F \ll 1$), whereas the second order approximation remains a very close approximation of the exact solution for both subcritical and supercritical conditions and for all Froude numbers.

**Acknowledgements**

This study is partially supported by the National Science Foundation (Award Numbers 1736713).

**Data Availability Statement**

No data is used or generated in this study.

## References

Aberson, S. D., J. A. Zhang and K. N. Ocasio (2017). "An Extreme Event in the Eyewall of Hurricane Felix on 2 September 2007." Monthly Weather Review **145**(6): 2083-2092.

Conner, W. C., R. H. Kraft, and D. L. Harris (1957). "Empirical Methods for Forecasting the Maximum Storm Tide Due to Hurricanes and Other Tropical Storms." *Monthly Weather Review* **85** (4): 113–16.

Folmer, M. J., M. DeMaria, R. Ferraro, J. Beven, M. Brennan, J. Daniels, R. Kuligowski, H. Meng, S. Rudlosky, L. Zhao, J. Knaff, S. Kusselson, S. D. Miller, T. J. Schmit, C. Velden and B. Zavodsky (2015). "Satellite tools to monitor and predict Hurricane Sandy (2012): Current and emerging products." Atmospheric Research **166**: 165-181.

Han, G., Z. Ma, N. Chen, N. Chen, J. Yang and D. Chen (2017). "Hurricane Isaac storm surges off Florida observed by Jason-1 and Jason-2 satellite altimeters." Remote Sensing of Environment **198**: 244-253.

Harris, D. L. (1959). "An Interim Hurricane Storm Surge Forecasting Guide," National Hurricane Research Project. Report No. 32, U.S. Weather Bureau, 1959, 24 PP.

Hoover, R. A. (1957). "Empirical Relationships of the Central Pressures in Hurricanes to the Maximum Surge and Storm Tide." *Monthly Weather Review* 85 (5): 167–74.

Irish, J. L., D. T. Resio, and J. J. Ratcliff (2008), The influence of storm size on hurricane surge, J. Phys. Oceanogr., 38, 2003– 2013.

Jelesnianski, C. P. 1966. "Numerical Computations of Storm Surges without Bottom Stress." *Monthly Weather Review* 94 (6): 379–94. https://doi.org/10.1175/1520-0493(1966)094<0379:NCOSSW>2.3.CO;2.

Hasler, A. F., Morris, K.R. (1986). "Hurricane structure and wind fields from stereoscopic and infrared satellite observations and radar data " J. Climate appl. Met., **25**(6): 709–727.

Pellikka, H., J. Šepić, I. Lehtonen and I. Vilibić (2022). "Meteotsunamis in the northern Baltic Sea and their relation to synoptic patterns." Weather and Climate Extremes **38**: 100527.

Layek, G. C., & Pati, N. C. (2017). Bifurcations and chaos in convection taking non-Fourier heat-flux. *Physics Letters A*, *381*(41), 3568-3575. https://doi.org/https://doi.org/10.1016/j.physleta.2017.09.020.

Peng, M., L. Xie, and L. J. Pietrafesa (2004), A numerical study of storm surge and inundation in the Croatan-Albemarle-Pamlico estuary system, Estuarine Coastal Shelf Sci., 59, 121–137.

Peng, M., L. Xie, and L. J. Pietrafesa (2006a), A numerical study on hurricane-induced storm surge and inundation in Charleston Harbor, South Carolina, J. Geophys. Res., 111, C08017, doi:10.1029/ 2004JC002755.

Peng, M., L. Xie, and L. J. Pietrafesa (2006b), Tropical cyclone induced asymmetry of sea level surge and fall and its presentation in a storm surge model with parametric wind fields, Ocean Modell., 14, 81– 101.

Proudman, J. (1929). "The fffects on the sea of changes in atmospheric pressure." Geophys. J. Int. **2**: 197-209.

Proudman, J. (1953), Dynamical Oceanography, 409 pp., John Wiley, New York.

Rego, J. L. and C. Li (2009). "On the importance of the forward speed of hurricanes in storm surge forecasting: A numerical study." *Geophysical Research Letters* **36**(7): L07609 (07601-07605).

Saffir, H. S. (1973). "Hurricane Wind and Storm Surge." *The Military Engineer* **65**(423): 4-5.

Šepić, J., I. Vilibić, and D. Belusic (2009), Source of the 2007 Ist meteotsunami (Adriatic Sea), J. Geophys. Res., 114, C03016, doi:10.1029/2008JC005092.

Shao, M., Zhang, M., Wu, J., Guo, X., Wu, Q., & Qing, J. (2023). Bifurcation and chaos analysis of a pretensioned moving printed electronic laminated membrane considering aerothermoelasticity. *Results in Physics*, *44*, 106148. https://doi.org/https://doi.org/10.1016/j.rinp.2022.106148.

Shen, F., D. Xu, J. Min, Z. Chu and X. Li (2020). "Assimilation of radar radial velocity data with the WRF hybrid 4DEnVar system for the prediction of hurricane Ike (2008)." Atmospheric Research **234**: 104771.

Simpson, R. H. (1974). "The Hurricane Disaster Potential Scale." *Weatherwise* **27**(169): 169-186.

Sheremet, A., U. Gravois and V. Shrira (2016). "Observations of meteotsunami on the Louisiana shelf: a lone soliton with a soliton pack." Natural Hazards **84**: 471-492.

Yang, L., J. Smith, M. Liu and M. L. Baeck (2019). "Extreme rainfall from Hurricane Harvey (2017): Empirical intercomparisons of WRF simulations and polarimetric radar fields." Atmospheric Research **223**: 114-131.

Zhang, C., and C. Li. Effects of Hurricane Forward Speed and Approach Angle on Storm Surges: An Idealized Numerical Experiment, *Acta Oceanologica Sinica,* **38**(7): 1-9, DOI:10.1007/s13131-018-1081-z.

Zhu, P., J. A. Zhang and F. D. Marks (2023). "On the Lateral Entrainment Instability in the Inner Core Region of Tropical Cyclones." Geophysical Research Letters **50**(8): e2022GL102494.